# FABRICATION OF HIGH QUALITY QC FILMS VIA THE ROUTE OF THE AMORPHOUS PHASE

R. HABERKERN, C. ROTH, R. KNÖFLER, S. SCHULZE, P. HÄUSSLER
Inst. f. Physik, TU Chemnitz, 09107 Chemnitz, Germany, r.haberkern@physik.tu-chemnitz.de

## ABSTRACT

We discuss the preparation of thin icosahedral films (Al-Cu-Fe and Al-Pd-Re) via the route of the amorphous (*a*-)phase which in some aspects is a precursor to the icosahedral phase. A direct transition from the a- to the *i*-phase occurs for Al-Cu-Fe films at 430°C on the time scale of minutes. The resulting films are of good quality as shown by diffraction and electronic transport properties. The surface of the resulting films is very smooth.

## INTRODUCTION

Most of the thermodynamically stable quasicrystals show properties which are not usual for Al-based alloys. Especially mechanical properties like the huge hardness and a small friction coefficient might be of technical interest as they are combined with a high chemical stability [1,2]. But at not too elevated temperatures quasicrystals are brittle. Therefore, applications of quasicrystals as thin films or coatings will be much more likely than the use of quasicrystals as bulk materials. Some efforts have been made to produce such films by different methods like plasma spraying or PVD methods [3].

We want to introduce a method of preparing thin quasicrystalline (qc) films by a short term annealing of the amorphous phase which is in some aspects a precursor to quasicrystals [4]. The resulting films are characterized by electron diffraction and electronic transport properties which are very sensitive to the interaction between the electronic system and the peculiar atomic structure of quasicrystals [5]. The occurrence of large anomalies in transport properties corresponds to the transition of nearly all qc properties from a nearly metallic to a more covalent behaviour.

## FABRICATION OF AMORPHOUS FILMS

Films considered here were produced by vapour quenching to a substrate (quartzglass, sapphire or silicon) held at room temperature either by sequential flash evaporation (Al-Cu-Fe) or by a sputtering technique (Al-Pd-Re).

The sequential flash evaporation technique uses *one* vapour source, although the vapour pressure of the elements may differ by orders of magnitude. The source mainly consists of the hot tungsten filament shaped as seen in Fig. 1 with two edges in order to get a homogeneous temperature distribution in the curved region, and a tube, tapped inside, which allows, due to rotation, the feeding of small pieces of the material onto the filament. Each piece has to have the overall nominal composition and gets flash evaporated. As each contributes one monolayer or less to the film thickness, even a complete segregation of each piece during evaporation finally ends up with a homogeneous sample of the nominal composition [6]. The master alloy of Al-Cu-Fe has been mixed by arc-melting followed by crunching to small grains with a typical size of about 100 µm.

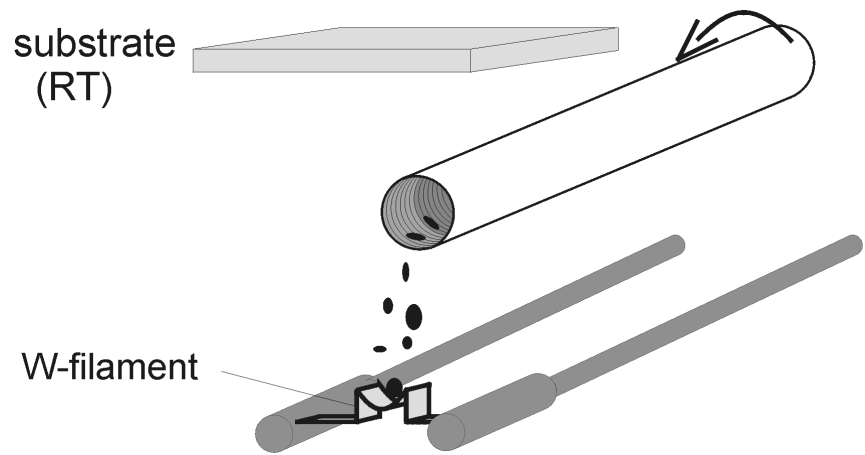

Fig. 1:   Sketch of the sequential flash-evaporation device

As a second technique for Al-Pd-Re we use co-sputtering with two magnetron sources. Due to the positions of the two sources in respect to the substrate a defined composition gradient can be achieved along the substrate. In one preparation process a set of amorphous samples can be produced consisting of about 20 samples with a composition, slightly and systematically changing from one sample to the next, cutting the ternary phase diagram at, or close to the optimal composition. A defined shape of the evaporated or sputtered samples can be achieved by applying a mask or a structuring technique using microelectronic technologies. This allows to determine e.g. the absolute resistivity, which is difficult for most bulk samples because of their irregular shape.

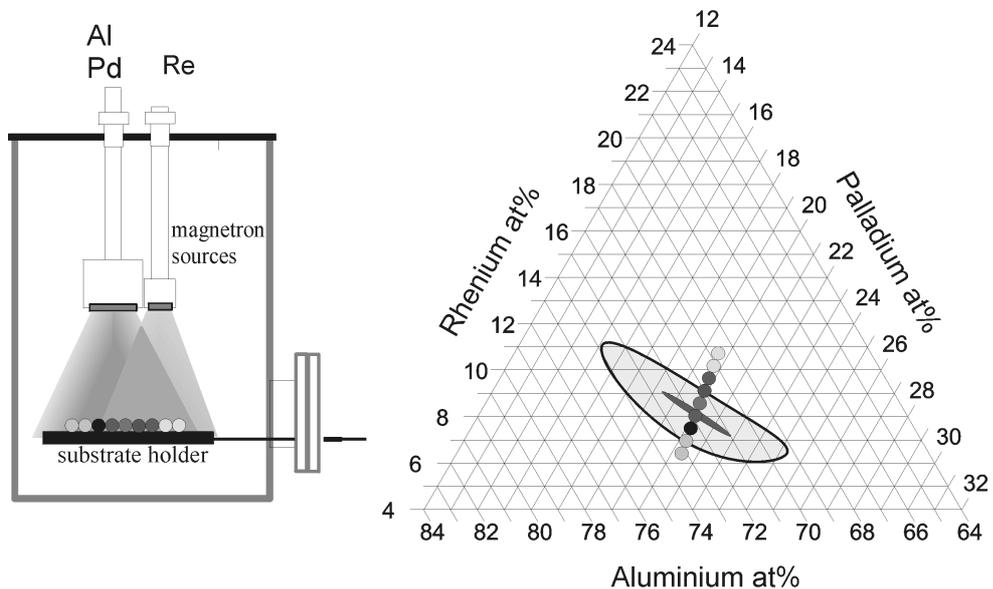

Fig. 2:   Sketch of the co-sputtering device (left) and a part of the ternary phase diagram (right) showing the presumed area of the i-phase (shadowed) and a sequence of samples co-sputtered in one deposition step

As film thickness a value of typically 50 – 300 nm was used in case of flash evaporation and of 200 nm in case of co-sputtered films. These values are well inside a 3-dimensional regime with respect to the relevant inelastic scattering length, at least down to $T$=0.1K.

# AMORPHOUS TO QUASICRYSTALLINE TRANSITION

The amorphous state prepared by quenching of the vapor phase onto a substrate held at room temperature or cooled to 4K is structurally similar to the liquid phase. Therefore, a direct transition from the amorphous to the qc-phase is possible for films at appropriate composition. In contrast to the transition from the liquid to the quasicrystal, the transition from the amorphous to the qc-state is irreversible and occurs at much lower temperatures.

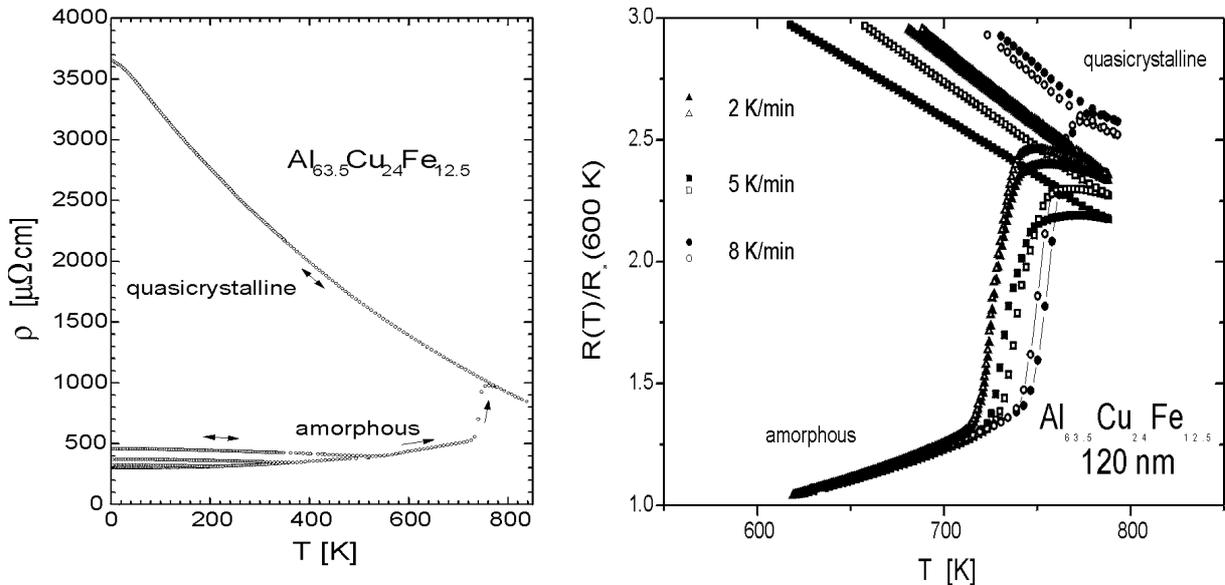

Fig. 3: Electrical resistivity (left) of a thin film of Al-Cu-Fe prepared at $T=4$ K in the amorphous state. After some annealing steps in the amorphous sample, the crystallization to the icosahedral phase occurs at $T=740$ K. On the right the transition is plotted enlarged for different heating rates

Fig. 3 shows the resistivity of a thin film of Al-Cu-Fe deposited by sequential flash evaporation onto a substrate cooled down to $T=4$K. The resistivity of the amorphous state at $T=4$K is $\rho=300$ μΩcm and increases to a value of $\rho=470$ μΩcm after an annealing up to 550K. Besides this quite small and continuous irreversible increase of the resistivity and its temperature coefficient (TCR), a distinct increase can be seen at $T\approx740$ K. Above, the sample shows a large negative TCR which is typical for structurally good quasicrystals. Electron diffraction shows the icosahedral symmetries as reproduced in Fig. 4 for a free standing film without any ion-etching or similar techniques.

The transition temperature is about 350 K lower than the solidus temperature in the same system. This offers new possibilities. First, the transition happens within the solid phase and allows the measurement of electronic transport properties at the transition, on a time scale which is reduced by many orders of magnitude due to the low temperature compared to the liquid to qc transition. Second, the transition plotted in fig. 3 shows no indication for the formation of crystalline phases like the cubic β-phase which occur whenever a Al-Cu-Fe melt is rapidly cooled down from the liquid state by melt-spinning. Therefore, in films produced via the amorphous route no long annealing treatments (typically hours at $T=1050$ K for bulk samples) are necessary for removing crystalline phases. Third, the transition from the amorphous to the qc phase happens not only at low temperatures but also on short time scales of the order of minutes. In fig. 3 (right) transitions of $Al_{63.5}Cu_{24}Fe_{12.5}$ are drawn for different heating rates, which indicates a transition time between 3 and 15 minutes according to temperatures between 750 and 720 K, respectively. This may open the possibility of a flash annealing without a relevant heating of the substrate. Together with the low surface roughness achieved by this preparation (Fig. 6) this may have consequences for technical applications of qc coatings.

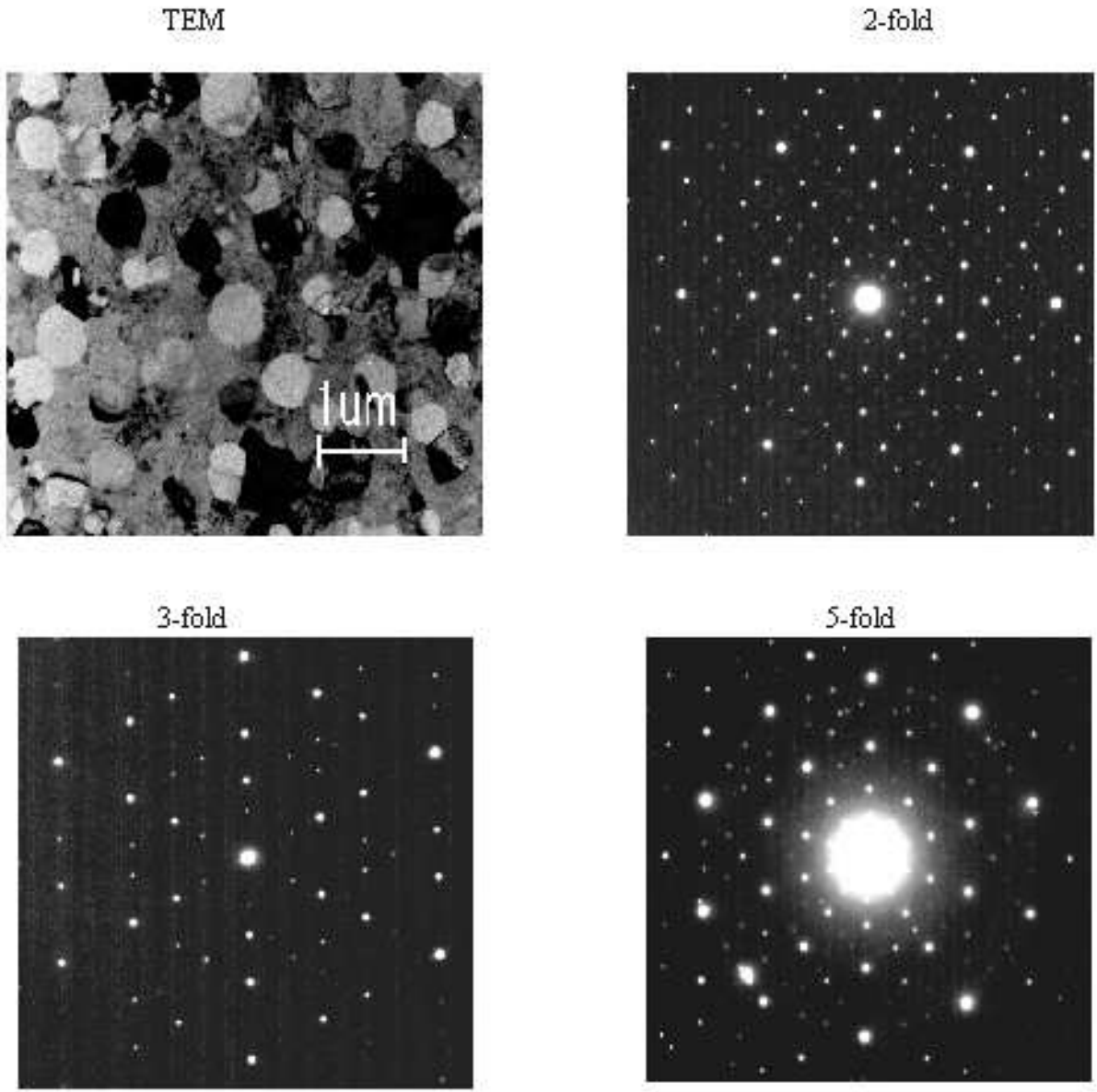

Fig. 4: Electron diffraction of a free carrying i-Al-Cu-Fe film of thickness $d$=120 nm

It has been shown that electronic transport properties are very sensitive to the structure quality [7]. Especially the temperature coefficient of the electrical conductivity is widely used to characterize samples. Fig. 7 shows the conductivity $\sigma(T)$ of typical *i*-Al-Pd-Re and *i*-Al-Cu-Fe films. The $\sigma_{300K}/\sigma_{T\to 0}$ value for *i* Al-Cu-Fe films is typically between 1.6 and 1.9 and is not very different compared to a values between 1.9 and 2.4 reached for typical icosahedral bulk samples. For *i*-Al-Pd-Re films the $\sigma_{300K}/\sigma_{T\rightarrow 0}$ ratio tends to infinity as well as for bulk samples. We could show that also films are insulating, that means their $T\rightarrow 0$ conductivity tends to $\sigma=0$ as determined from the conductivity behaviour for temperatures down to about 100 mK [8]. The conductivity for $70K < T < 900K$ could be attributed to the resonant interaction between the electronic and the atomic system due to a simple function of the Debye-Waller factor as shown by the fit in Fig.7 (left) and discussed in [9].

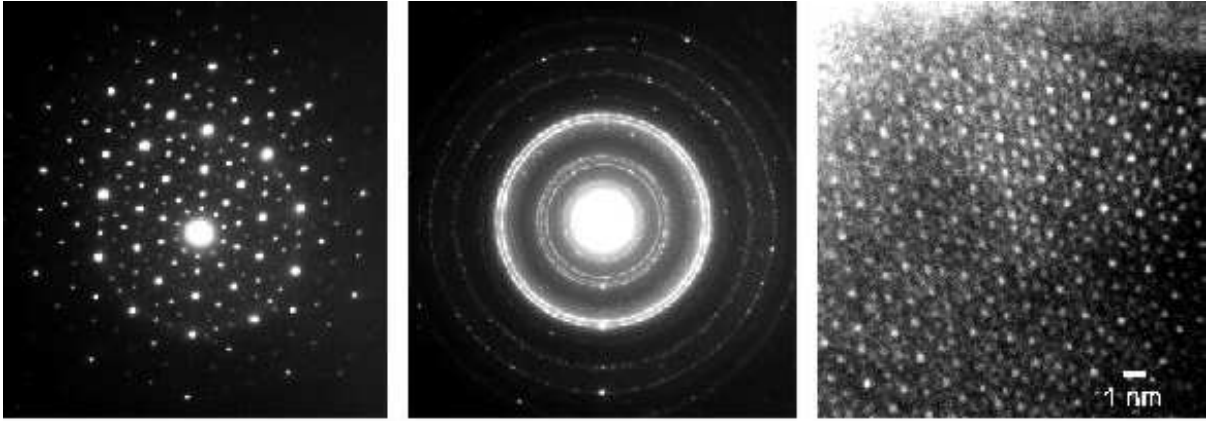

Fig.5: Electron diffraction on a thinned Al-Pd-Re film, along a five-fold axis (left), large area electron diffraction (middle) and HREM (right)

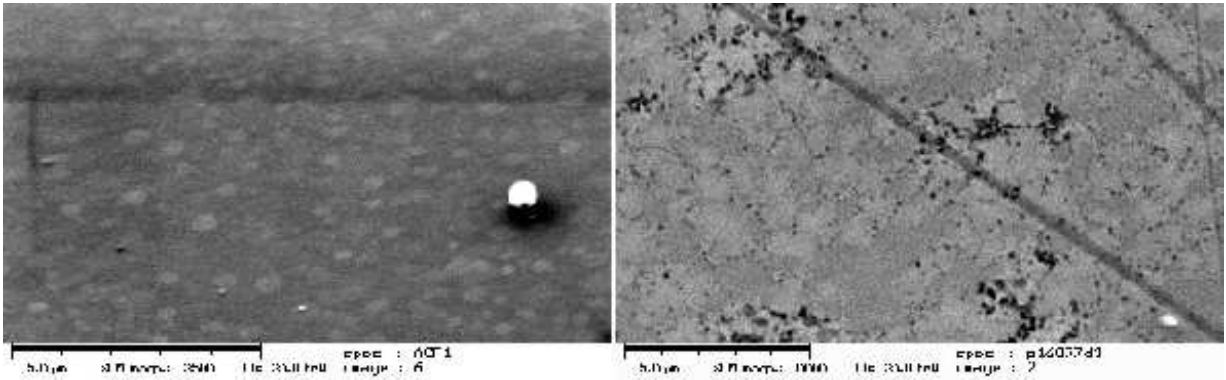

Fig. 6: SEM of an *i*-Al-Cu-Fe (left) and an *i*–Al-Pd-Re film (right). Black areas in the Al-Pd-Re film correspond mostly to holes

**CONCLUSIONS**

The preparation via the route of the amorphous phase provides a method of getting *i*-films of high quality and with a smooth surface by a short term annealing treatment at quite low temperatures. Film quality has been probed by electron diffraction, HREM and electronic transport measurements. The *i*-Al-Pd-Re films introduced here are the first which show large transport anomalies up to a metal-insulator transition.

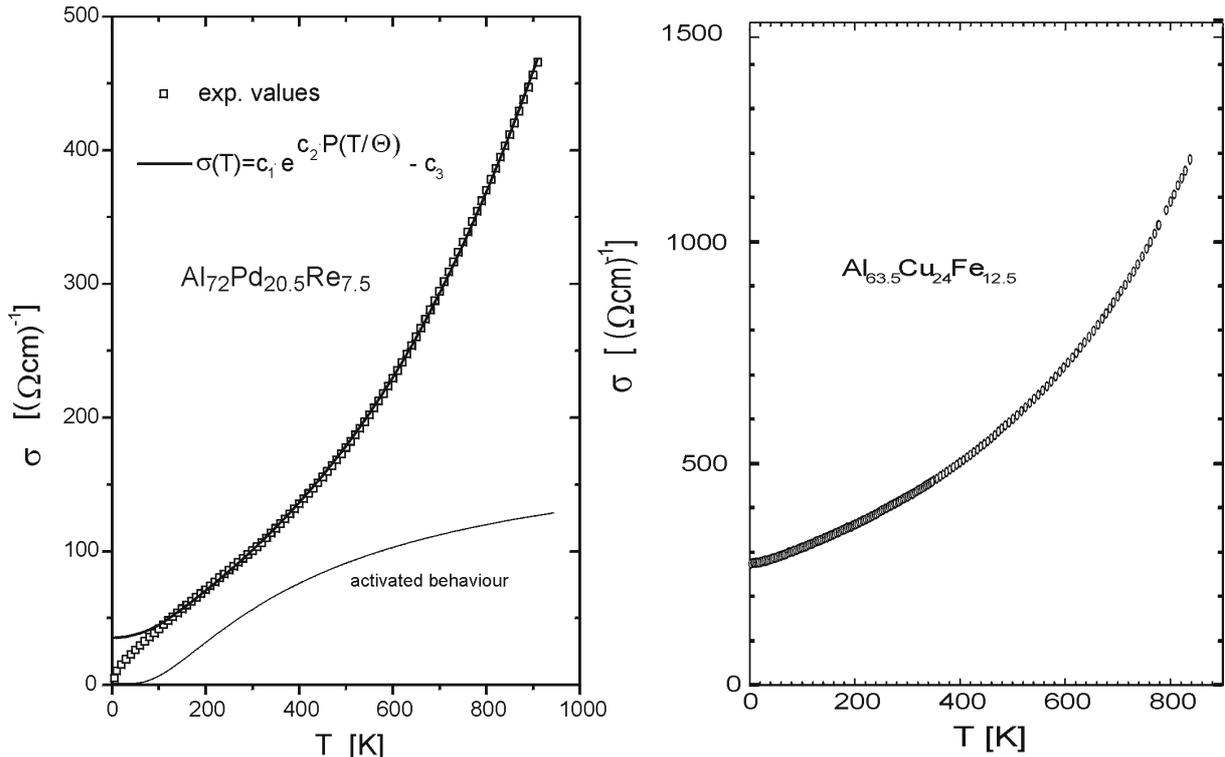

Fig. 7: Conductivity of an *i*-Al-Pd-Re film (left) and an *i*-Al-Cu-Fe film (right) as a function of $T$

## ACKNOWLEDGEMENTS

This work was supported by Deutsche Forschungsgemeinschaft